\documentclass[11pt]{article}

\usepackage{epsfig}
\usepackage{subfig}
\usepackage{graphicx}
\usepackage{epstopdf}
\usepackage{amsfonts}
\usepackage{amssymb}
\usepackage{amsthm}
\usepackage{amsmath}
\usepackage{multirow}
\usepackage{color}
\usepackage{cite}
\usepackage{xcolor}
\usepackage{makecell}
\def\beq{\begin{equation}}
\def\eeq{\end{equation}}
\def\bea{\begin{eqnarray}}
\def\eea{\end{eqnarray}}
\newcommand{\beqs}{\begin{subequations}}
\newcommand{\eeqs}{\end{subequations}}

\newcommand{\cref}[1]{Ref.~\cite{#1}}

\newcommand{\hh}{{\ensuremath{I{\kern-2.6pt h}}}}
\newcommand{\bhh}{{\ensuremath{\bar{I{\kern-2.6pt h}}}}}

\usepackage[colorlinks=true,allcolors=blue]{hyperref}

\begin{document}

\begin{titlepage}
	

\begin{center}
{\Large {\bf Quantum tunneling in the early universe:\\ Stable magnetic monopoles from metastable cosmic strings}
}
\\[12mm]
George Lazarides,$^{1}$
Rinku Maji,$^{2,3}$
Qaisar Shafi$^{4}$~
\end{center}
\vspace*{0.50cm}
\centerline{$^{1}$ \it
School of Electrical and
Computer Engineering, Faculty of Engineering,
}

\centerline{\it
Aristotle University
of Thessaloniki, Thessaloniki 54124, Greece}
\vspace*{0.2cm}
	\centerline{$^{2}$ \it
		Laboratory for Symmetry and Structure of the Universe, Department of Physics,}
		\centerline{\it  Jeonbuk National University, Jeonju 54896, Republic of Korea}
	\vspace*{0.2cm}
	\centerline{$^{3}$ \it
		Cosmology, Gravity and Astroparticle Physics Group, Center for Theoretical Physics of the Universe,}
		\centerline{\it  Institute for Basic Science, Daejeon 34126, Republic of Korea}
	\vspace*{0.2cm}
	\centerline{$^{4}$ \it
		Bartol Research Institute, Department of Physics and 
		Astronomy,}
	\centerline{\it
		 University of Delaware, Newark, DE 19716, USA}
	\vspace*{1.20cm}
\begin{abstract}
\noindent We present a novel mechanism for producing topologically stable monopoles (TSMs) from the quantum mechanical decay of metastable cosmic strings in the early universe. In an $SO(10)$ model this mechanism yields TSMs that carry two units ($4\pi/e$) of Dirac magnetic charge as well as some color magnetic charge which is screened. For a dimensionless string tension parameter $G\mu \approx 10^{-9} - 10^{-5}$, the monopoles are superheavy with masses of order $10^{15} - 10^{17}$ GeV. Monopoles with masses of order $10^8 - 10^{14}$ GeV arise from metastable strings for $G\mu$ values from $\sim 10^{-22}$ to $10^{-10}$. We identify the parameter space for producing these monopoles at an observable level with detectors such as IceCube and KM3NeT. For lower $G\mu$ values the ultra-relativistic monopoles should be detectable at Pierre Auger and ANITA. The stochastic gravitational wave emission arise from metastable strings with $G\mu\sim 10^{-9}-10^{-5}$ and should be accessible at HLVK and future detectors including the Einstein Telescope and Cosmic Explorer. An $E_6$ extension based on this framework would yield TSMs from the quantum mechanical decay of metastable strings that carry three units ($6\pi/e$) of Dirac magnetic charge.
\end{abstract}

\end{titlepage}
\section{Introduction}
Unified theories based on $SU(4)_c \times SU(2)_L \times SU(2)_R$ \cite{Pati:1974yy} and $SO(10)$ \cite{Georgi:1974my,Fritzsch:1974nn} provide a compelling extension of the Standard Model (SM). In addition to electric charge quantization these theories predict the existence of right-handed neutrinos and non-zero Majorana masses for the SM neutrinos as well as neutrino oscillations\cite{Lazarides:1980nt}. The presence of electric charge quantization implies the existence of a superheavy 't Hooft-Polyakov type monopole \cite{tHooft:1974kcl,Polyakov:1974ek} carrying one unit of Dirac magnetic charge \cite{Dirac:1931kp} and some color magnetic charge \cite{Daniel:1979yz, Dokos:1979vu, Lazarides:1982jq}. Depending on the symmetry breaking pattern, theories based on $SO(10)$ can yield a doubly charged monopole that is lighter than the singly charged monopole mentioned above \cite{Lazarides:1980cc}. $SO(10)$ grand unification also predicts the existence of topologically stable cosmic strings if the symmetry breaking to $SU(3)_c \times U(1)_{em}$ is implemented using only tensor representations \cite{Kibble:1982ae}. The dimensionless string tension parameter $G \mu$ for such strings, where $G$ and $\mu$ respectively denote Newton’s gravitational constant and the energy per unit length of the string,  is determined by the underlying symmetry breaking pattern of $SO(10)$. Compatibility with the recent Pulsar Timing Array (PTA) experiments \cite{NANOGrav:2023gor,NANOGrav:2023hvm,Antoniadis:2023ott,Reardon:2023gzh,Xu:2023wog} and the LIGO O3 bound \cite{LIGOScientific:2021nrg} require that $G\mu \lesssim 10^{-10}$ and $G\mu \lesssim 10^{-7}$ respectively for these topologically stable cosmic strings.  

$SO(10)$ breaking can also yield metastable strings \cite{Preskill:1992ck,Martin:1996ea,Martin:1996cp,Leblond:2009fq,Buchmuller:2019gfy,Buchmuller:2020lbh,Blasi:2020wpy,Masoud:2021prr,Dunsky:2021tih,Ahmed:2022rwy,Afzal:2022vjx,Buchmuller:2021dtt,Buchmuller:2021mbb,Chitose:2023dam}, which produce a stochastic gravitational background that appears compatible with the latest PTA data for $G \mu \approx 10^{-7}$ and the metastability factor $\sqrt{\kappa}\approx 8$ \cite{Lazarides:2023ksx,Antusch:2023zjk,Lazarides:2023rqf,Maji:2023fhv,Ahmed:2023rky,Fu:2023mdu,Afzal:2023cyp,Afzal:2023kqs,Ahmed:2023pjl,Lazarides:2023bjd,Pallis:2024mip}. Here, $\kappa = m_M^2 / \mu$, with $m_M$ denoting the monopole (and antimonopole) mass whose presence makes the cosmic strings metastable through quantum mechanical tunneling.

Yet another cosmic string explanation for the stochastic gravitational radiation observed by the PTA experiments is closely related to the metastable string scenario. This so-called quasistable string (QSS) scenario \cite{Lazarides:2022jgr} also employs superheavy cosmic strings with the metastability factor $\sqrt{\kappa}\gtrsim 9$. The cosmic string decay via quantum tunneling in the metastable case is replaced in QSS by the reappearance in the particle horizon of the primordial monopoles and antimonopoles that undergo only partial inflation.

In this paper we explore a novel cosmic string scenario in which $SO(10)$ symmetry breaking yields metastable strings whose subsequent decay not only produces a stochastic gravitational background but also yields an observable number density of topologically stable monopoles (TSMs). The quantum mechanical tunneling of two distinct varieties of monopoles and antimonopoles on the strings yields, through their coalescing, the corresponding stable monopoles and antimonopoles. In the example we discuss in this paper the stable monopole from the string decay carries two units ($4 \pi / e$) of Dirac magnetic charge as well as some color magnetic charge \cite{Lazarides:2019xai}. It appears from the merger of two confined monopoles on the string carrying magnetic charges ($8 \pi /3 e$)  and ($4 \pi /3e$) respectively. This doubly charged monopole is stable because the single charged Dirac monopole is produced at the first stage of $SO(10)$ breaking and is therefore heavier. The strings being metastable also produce a stochastic gravitational background spectrum which, for $\sqrt{\kappa}\approx 6$, displays a peak not far below the current LIGO O3 bound. However, the stable monopole abundance is quite suppressed in this case. On the other hand, the number density of the stable monopoles from the decay of strings after the end of the friction dominated era lies within reach of the ongoing and future experiments for $G \mu \sim 10^{-5} - 10^{-9}$ and $\sqrt{\kappa}$ around $5$, but the gravitational wave spectrum may be challenging to detect.

In Section \ref{sec:2}, we describe the formation of TSM from the merger of an $SU(4)_c$ monopole and an $SU(2)_R$ monopole in $SO(10)$. Section \ref{sec:em-rad} discusses the radiation of massless gauge bosons from the monopoles with unconfined magnetic fluxes at the end points of a string segment before the electroweak breaking. We estimate the present day TSM number density from metastable strings and the observational constraints on the present monopole flux in Section \ref{sec:4}. In Section \ref{sec:gws}, we discuss the high frequency gravitational waves from metastable strings and their possible observational prospects. In Section \ref{sec:fric-era}, we provide a rough estimate of the stable monopole number density from metastable strings that decay during the friction dominated era. Our conclusions are summarized in Section \ref{sec:conc}.
\section{SO(10) symmetry breaking, monopoles and metastable string}
\label{sec:2}
Consider the following $SO(10)$ symmetry breaking to the SM gauge group $\mathcal{G}_{\rm SM}$:
\begin{align}\label{eq:breaking-chain}
SO(10) \to & {SU(4)_c \times SU(2)_L \times SU(2)_R}/{Z_2} \nonumber\\ \to & {SU(3)_c \times U(1)_{B-L} \times SU(2)_L \times U(1)_R}/{Z_3\times Z_2} \nonumber\\ \to & {SU(3)_c  \times SU(2)_L \times U(1)_Y}/{Z_3\times Z_2} .
\end{align}     
\begin{figure}[ht!]
\begin{center}
\includegraphics[width=0.6\textwidth]{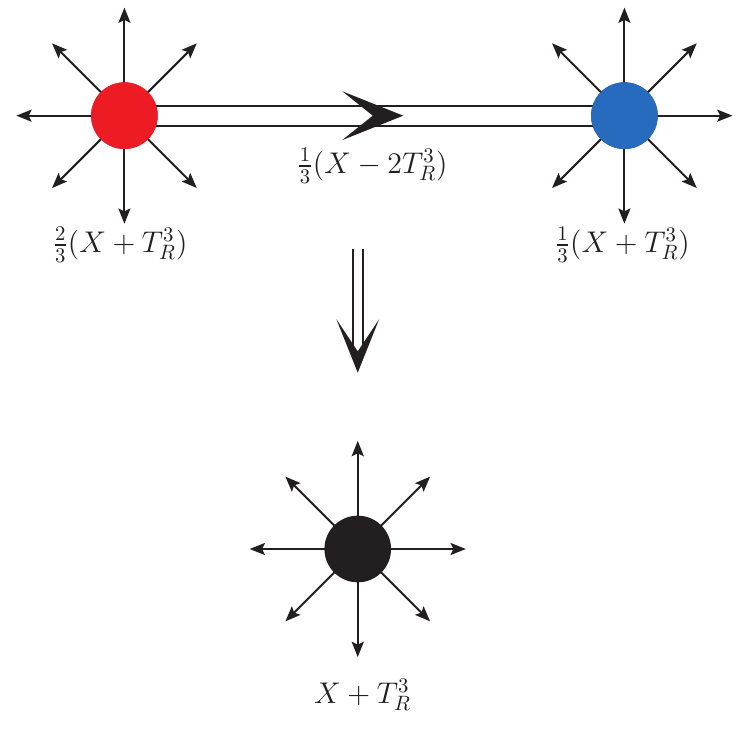}
\end{center}
\caption{Formation of the topologically stable monopole from the merger of a red ($SU(4)_c$) and a blue ($SU(2)_R$) monopole connected by a string segment before the 
EW symmetry breaking.}\label{fig:monopole_merge_1}
\end{figure}  
\begin{figure}[ht]
\begin{center}
\includegraphics[width=0.6\textwidth]{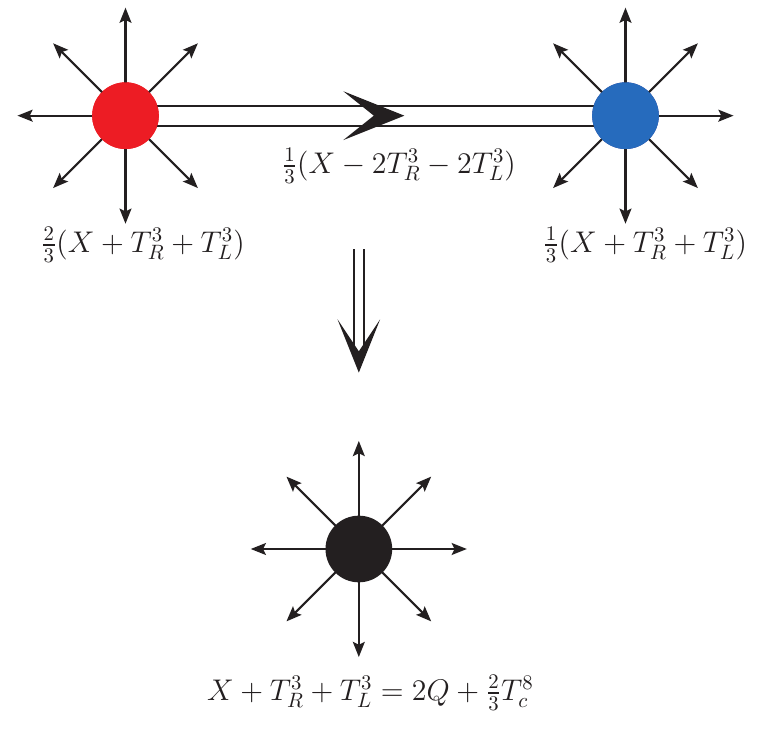}
\end{center}
\caption{Formation of the stable monopole from the merger of a red and a blue monopole after the EW breaking. This formation process can also 
appear alternatively to the one in Fig. 1 even if the merger occurs before the EW breaking. The stable monopole carries two quanta ($4\pi/e$) of Dirac magnetic charge as well as color magnetic charge. Note that the red and blue monopoles carry electromagnetic magnetic flux corresponding to their magnetic charge ($8\pi/3e$) and ($4\pi/3e$) respectively.}\label{fig:monopole_merge_2}
\end{figure}
\hspace{-2mm}The first breaking produces a topologically stable monopole that subsequently evolves into the superheavy GUT monopole that carries a single unit of Dirac magnetic charge ($2 \pi / e$) as well as color magnetic charge. From the second breaking we obtain two sets of monopoles, namely an 
$SU(4)_c$ monopole and an $SU(2)_R$ monopole with Coulomb magnetic 
fluxes $X=B-L+2T^8_c/3$ and $T^3_R$ respectively, where $T^8_c=diag(1, 1, 
-2)$ and $T^3_R=diag(1,-1)$.
Following the nomenclature in Ref.~\cite{Lazarides:2019xai}, we refer to these monopoles as the `red’ and `blue’ monopole respectively. In the third step in Eq.~\eqref{eq:breaking-chain} the symmetry $U(1)_{B-L} \times U(1)_R$ is spontaneously broken to $U(1)_Y$  by the vacuum expectation value (VEV) of a Higgs field $\nu^c_H$ with the quantum numbers of the right-handed neutrinos $\nu^c$. This generates a string that ends up confining the red and blue monopoles with their respective antimonopoles.

However, as shown in Ref.~\cite{Lazarides:2019xai}, this string can also connect a red and a blue monopole causing their merger which yields an unconfined monopole that carries two quanta $(4\pi/e)$ of Dirac magnetic charge as well as color magnetic charge. The existence of such a monopole is expected on topological grounds in the symmetry breaking pattern in 
Eq.~\eqref{eq:breaking-chain}. A dumbbell-like configuration depicting this structure and the emergence of the stable doubly charged monopole before the electroweak (EW) breaking is shown in Fig.~\ref{fig:monopole_merge_1}.

At this stage $SU(2)_L$ is unbroken and, thus, a $2\pi$ rotation around $T^3_L=diag(1, -1)$ is contractable. Consequently, we can freely add a 
$T^3_L$ Coulomb magnetic flux to the blue monopole and rearrange the fluxes so that the unconfined magnetic fields of the monopoles are 
along the $X+T^3_R+T^3_L$ direction, which remains unbroken even after the EW breaking. The resulting configuration is shown in Fig.~\ref{fig:monopole_merge_2} and remains unaltered by the EW breaking. Note that the electromagnetic magnetic flux of the red and blue monopoles in this configuration corresponds to their unconfined magnetic charge $(8\pi/3e)$ and $(4\pi/3e)$ respectively. If the merger of a red and a blue monopole takes place before the EW breaking via the structure in Fig.~\ref{fig:monopole_merge_1}, we can add the $T^3_L$ flux directly to the resulting stable monopole so that it remains unaffected by the EW breaking.

As we go around the flux tubes in Figs.~\ref{fig:monopole_merge_1} and \ref{fig:monopole_merge_2}, the phase of the $\nu^c_H$ VEV changes by $2\pi$, which implies that these tubes carry $\nu^c$ zero modes. On the contrary, the phases of the EW Higgs doublet VEVs remain constant around the string in Fig.~\ref{fig:monopole_merge_2} and, thus, this string does not carry charged fermionic zero modes, i.e., it is not superconducting. Finally, it is worth mentioning that the strings connecting a red monopole to its antimonopole or a blue antimonopole to its monopole 
are the same as the strings depicted in Figs.~\ref{fig:monopole_merge_1} and \ref{fig:monopole_merge_2}.

We follow here the standard metastable string scenario according to which the primordial red and blue monopoles together with their antimonopoles are inflated away. Referring to Eq.~\eqref{eq:breaking-chain} and the discussion above, the GUT monopole is also inflated away, and the breaking of the symmetry $U(1)_{B-L} \times U(1)_R$ to $U(1)_Y$ that yields the metastable cosmic string occurs after inflation (for an example of how this type of inflationary scenario can be realized, see Ref.~\cite{Lazarides:2008nx}).

Our main challenge now is to find out whether an observable flux of the stable monopole in Fig.~\ref{fig:monopole_merge_1} can result from the decay of the metastable cosmic string via the quantum mechanical tunneling of the red and blue monopoles and their antimonopoles.
\section{Radiation of massless gauge bosons from monopoles}
\label{sec:em-rad}
Consider a sub-horizon segment with a red and a blue monopole at its end points. It becomes almost straight by the expansion, oscillates and fragments. The acceleration of the two monopoles as they are pulled towards each other by the string segment leads to radiation of massless gauge bosons. The rate of energy loss from each of the two monopoles due to this radiation is \cite{Berezinsky:1997kd, Leblond:2009fq, Kibble:2015twa}
\begin{align}
\label{eq:dE/dt}
\frac{dE}{dt}=-\frac{1}{6\pi}g_m^2a^2 ,
\end{align}
where $g_m$ is its magnetic charge and $a$ its proper acceleration, i.e., the acceleration in its instantaneous rest frame, which turns out to be $a = \mu/m_M$.

In the region of the parameter space that we mainly consider in this paper, the EW symmetry is unbroken and, thus, both the red and blue monopoles carry unconfined magnetic flux along the hypercharge direction $Y=\left[(B-L)+T^3_R\right]/2$ plus some colour magnetic flux, as one can deduce from Fig.~\ref{fig:monopole_merge_1}.
Since the chromomagnetic fields are screened, we only consider the unconfined $Y$ magnetic flux of the red and blue monopoles \cite{Lazarides:2023iim}
\begin{align}
\label{eq:mag-charge}
g_{mR} = \frac{2}{3}\left(\frac{4\pi}{g_1}\right)\quad \mathrm{and}\quad g_{mB} = \frac{1}{3}\left(\frac{4\pi}{g_1}\right) ,
\end{align}
where $g_1$ denotes the $U(1)_Y$ gauge coupling constant.
Thus, the rate of energy loss from both monopoles is given by
\begin{align}
\frac{dE}{dt} = -\frac{1}{6\pi}(g_{mR}^2+g_{mB}^2)a^2 = -\frac{10}{27\kappa}\frac{1}{\alpha_1(t)}\mu\equiv -j(t)\mu,
\end{align}
where $\alpha_1=g_1^2/4\pi$, $\kappa = m_M^2/\mu$ and $j=\frac{10}{27\kappa}\frac{1}{\alpha_1(t)}$.

We take as renormalization scale the temperature $T$ of the universe, 
which in the radiation dominated era is given by
\begin{align}
T=\left(\frac{45}{2\pi^2g_*}\right)^{\frac{1}{4}}\sqrt{\frac{m_{\rm Pl}}{t}},
\end{align}
where $g_*$ is the appropriate effective number of massless degrees of 
freedom and $m_{\rm Pl}$ is the reduced Planck scale. The one-loop renormalization group evolution of the gauge coupling constant $g_1$ is given by
\begin{align}\label{eq:RGE}
\frac{1}{\alpha_1(T)}=\frac{1}{\alpha_1(m_Z)}+\frac{b_1}{2\pi}\log\left(\frac{m_Z}{T}\right) ,
\end{align}
where $m_Z$ is the $Z$-boson mass and $b_1=41/6$ the one-loop $\beta$-coefficient \cite{Jones:1981we}.
 
 The length of the segment $l$ is defined as the total energy in the string divided by the string tension $\mu$. After including the radiation of gravitational wave, we can write
\begin{align}\label{eq:dl-dt}
\frac{dl}{dt} = -\tilde{\Gamma}G\mu-j(t) ,
\end{align}
where $\tilde{\Gamma}$ is the numerical factor for the radiation of gravitational waves from the string segments connecting the monopoles.

Note that in considering smaller values of $G\mu$, as we do in Section \ref{sec:fric-era}, a small region is encountered where the metastable strings decay with the EW symmetry broken, which happens for  $\log_{10}(G\mu) \lesssim \pi\kappa/2.3 - 63.7$. 
In this case the monopoles at the end points of a segment will radiate 
photons and the $U(1)_Y$ gauge coupling should be replaced by the 
electromagnetic one.
\section{Present monopole number density from metastable strings}
\label{sec:4}
The number density $\tilde{n}^{(s)}(l,t)$ of segments (per unit time per unit length) from long strings at cosmic time $t$ greater than the string decay time and length $l$ should satisfy the integro-differential Boltzmann equation \cite{Leblond:2009fq, Buchmuller:2021mbb}
\begin{align}\label{eq:n-seg}
\frac{\partial\tilde{n}^{(s)}(l,t)}{\partial t}+3H\tilde{n}^{(s)}(l,t)+\frac{\partial}{\partial l}\left(\frac{dl}{dt}\tilde{n}^{(s)}(l,t)\right) = \Gamma_d\left(2\int_l^\infty \tilde{n}^{(s)}(l',t)dl'- l\, \tilde{n}^{(s)}(l,t)\right),
\end{align}
where $H$ is the Hubble parameter, and $\Gamma_d = \frac{\mu}{2\pi}e^{-\pi\kappa}$ is the decay width per unit length of the metastable strings \cite{Preskill:1992ck}.
The solution of this equation with $dl/dt$ in Eq.~\eqref{eq:dl-dt} which matches with the scaling solution at the string decay time $t_s=\frac{1}{\sqrt{\Gamma_d}}$ is found to be 
\begin{align}\label{eq:sol-n-seg}
\tilde{n}^{(s)}(l,t) = \frac{4}{\xi^2}\frac{\Gamma_d^2}{4}\frac{(t+t_s)^2}{\sqrt{t^3t_s}}e^{-\Gamma_d\left[l(t+t_s)+\frac{1}{2}(\tilde{\Gamma}G\mu+j)(t-t_s)(t+3t_s)\right]},
\end{align}
where $\tilde{\Gamma}$ is taken about $50$ \cite{Buchmuller:2021mbb} and $\xi=0.27$ \cite{Buchmuller:2021mbb} in the radiation dominated universe. 

The number density of segments from string loops at $t>t_s$ can be approximated \cite{Buchmuller:2021mbb} by $\tilde{n}^{(l)} (l,t) = \sigma\tilde{n}^{(l,1)} (l,t)$, where $\sigma$ is a ``fudge" factor of order unity and $\tilde{n}^{(l,1)} (l,t)$ is the number density of the first generation of these segments which can be found by solving the Boltzmann equation
\begin{align}\label{eq:n-loop}
\frac{\partial\tilde{n}^{(l,1)}(l,t)}{\partial t}+3H\tilde{n}^{(l,1)}(l,t)+\frac{\partial}{\partial l}\left(\frac{dl}{dt}\tilde{n}^{(l,1)}(l,t)\right) = S(l,t).
\end{align}
Here
\begin{align}
S(l,t) = \Gamma_d l n(l,t)
\end{align}
is the source term for the first generation of segments from loops with
\begin{align}\label{eq:source-loop}
n(l,t) = \frac{0.18}{t^{3/2}(l+\Gamma G\mu t)^{5/2}}e^{-\Gamma_d\left[l(t-t_s)+\frac{1}{2}\Gamma G\mu(t-t_s)^2\right]}\Theta(0.18t_s-l-\Gamma G\mu (t-t_s))
\end{align}
being the number density of loops at time $t>t_s$. The numerical factor $\Gamma$ is taken \cite{Buchmuller:2021mbb} equal to $\tilde{\Gamma}=50$ and $\Theta$ is the Heaviside function. The solution of Eq.~\eqref{eq:n-loop} obtained by using the ``method of characteristics" at $t>t_s$ is \cite{Buchmuller:2021mbb}
\begin{align}
\label{eq:sol-n-loop}
\tilde{n}^{(l,1)}(l,t) = e^{-\int_{t_s}^tdt'\Gamma_d\bar{l}(t')}\int_{t_s}^tdt'\left(\frac{a(t')}{a(t)}\right)^3S(\bar{l}(t'),t') e^{\int_{t_s}^{t'}dt''\Gamma_d\bar{l}(t'')},
\end{align}
where $a(t)$ is the scale factor of the universe and $\bar{l}(t')$ the length at $t'$ of the segment with length $l$ at $t$. The latter can be found by solving Eq.~\eqref{eq:dl-dt} with the use of the one-loop renormalization group improved $U(1)_Y$ gauge coupling constant in Eq.~\eqref{eq:RGE} and turns out to be
\begin{align}
\bar{l}(t') = l + (\tilde{\Gamma}G\mu+j)(t-t')-\frac{10}{27\kappa}\frac{b_1}{4\pi}\left(t-t'(1+\log(t/t')\right).
\end{align}

\begin{figure}[ht]
\begin{center}
{\includegraphics[width=0.7\textwidth]{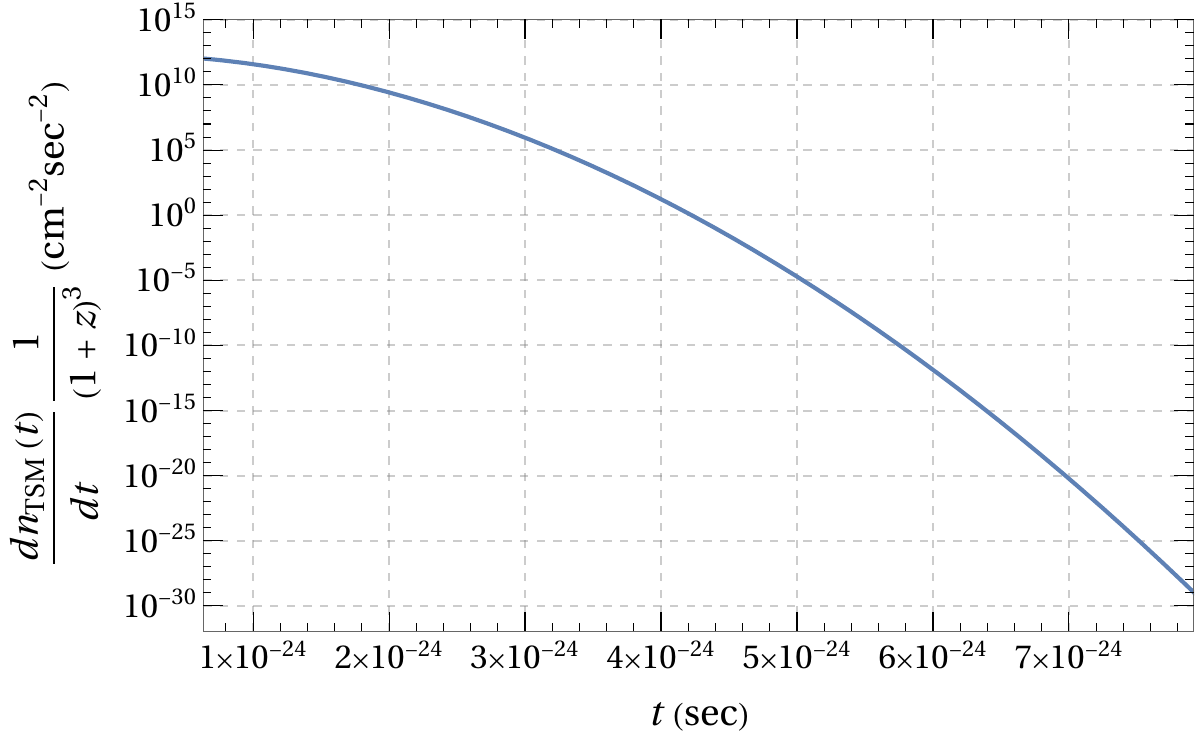}}
\end{center}
\caption{The rate of stable monopole formation per unit volume at cosmic time $t$ multiplied by the redshift factor for $G\mu = 10^{-6}$ and $\sqrt{\kappa}=4.8$ in units of $\mathrm{cm}^{-2}\mathrm{sec}^{-2}$.}
\label{fig:nM_t}
\end{figure}

The segments from both long strings and loops connect the red and blue monopoles to their respective antimonopoles, or the red monopoles (antimonopoles) to the blue monopoles (antimonopoles).
 Let us call the former as the `radiation segment' and the latter the $M_RSM_B$ structure. The strings with confined flux associated with $\frac{1}{3}(X-2T^3_R)$ can tunnel into the true vacuum via production of red or blue monopole-antimonopole ($M\bar{M}$) pair. There will appear a pair of radiation segments if a radiation segment connecting a red or blue monopole to its antimonopole gets cut into two pieces via the production of same type (red or blue) $M\bar{M}$, while two $M_RSM_B$ structures are produced via tunneling of opposite type (blue or red) of $M\bar{M}$ pair. An $M_RSM_B$ structure always produces an $M_RSM_B$ structure and one radiation segment.
 We can say that the $M_RSM_B$ structures generally constitute half of the total number of segments. The only exception is the first generation of segments sourced by string loops, which always connect red or blue monopoles to their respective antimonopoles. However we do not treat them separately since this difference can be absorbed in the fudge factor $\sigma$, which is expected to be small due to the strong energy loss to massless gauge boson radiation of the segments of all the generations. The $M_RSM_B$ structures with vanishing length ($l\to 0$) at cosmic time $t$ form TSMs at the same time. Therefore the rate of stable monopole formation per unit volume at time $t$ can be written as
\begin{align}
\frac{dn_{\rm TSM}(t)}{dt} = \frac{1}{2}\left(\tilde{n}^{(s)}(0,t) + \tilde{n}^{(l)}(0,t)\right) ,
\end{align}
and the present day monopole number density is 
\begin{align}
n_{\rm TSM} =\sum_{i\in \lbrace s,l\rbrace}n^{(i)}_{\rm TSM}=\frac{1}{2} \sum_{i\in \lbrace s,l\rbrace}\int_{t_s}^{t_e} dt\frac{ \tilde{n}^{(i)} (0,t)}{(1+z(t))^3},
\end{align}
where $n^{(s,l)}_{\rm TSM}$ are the monopole number densities from long strings and loops respectively, $t_e$ the time at which the monopole generation ceases to operate, and $z(t)$ the redshift.

Using Eqs.~\eqref{eq:sol-n-seg} and \eqref{eq:sol-n-loop}, we find that the contribution of segments from loops to the rate of stable monopole formation and their present number density is negligible compared to the contribution of segments from long strings for any reasonable value of the fudge factor $\sigma$ and, thus, we ignore it. In Fig.~\ref{fig:nM_t} we show the rate of formation of TSMs per unit volume at time $t$ multiplied by the redshift factor for $G\mu=10^{-6}$ and $\sqrt{\kappa}=4.8$.

The present stable monopole flux is given by $\mathcal{F}_{\rm TSM}=\frac{n_{\rm TSM}v_{\rm TSM}}{4\pi}$ in units of $\mathrm{cm}^{-2}\mathrm{sec}^{-1}\mathrm{sr}^{-1}$, with their present number density $n_{\rm TSM}$ in units of $\mathrm{cm}^{-2}\mathrm{sec}^{-1}$ and their mean local velocity $v_{\rm TSM}$ in units of the velocity of light.
\begin{table}[htbp!]
\begin{center}
\begin{tabular}{|c| c| c| c|}
\hline
\multirow{2}{*}{Experiment} & {Monopole mass ($m_{\rm TSM}$)} & \multirow{2}{*}{Velocity ($v_{\rm TSM}$)} & Flux $\mathcal{F}_{\rm TSM}= \frac{n_{\rm TSM}v_{\rm TSM}}{4\pi}$ \\
& in GeV && in $\mathrm{cm}^{-2}\mathrm{sec}^{-1}\mathrm{sr}^{-1}$ \\ 
\hline
\multirow{2}{*}{MACRO \cite{Ambrosio:2002qq}} & $5\times 10^8 - 5\times 10^{13}$ & $>0.05$ & $3\times 10^{-16}$ \\
\cline{2-4}
& $>5\times 10^{13}$  & $>4\times 10^{-5}$ & $1.4\times 10^{-16}$ \\
\hline
{IceCube \cite{IceCube:2021eye}} & $>10^{8}-10^{10}$ & $0.8 - 0.995$ & $2\times 10^{-19}$ \\
\hline
\end{tabular}
\caption{Upper limits on the monopole flux from the MACRO and IceCube experiments that the TSMs should obey \cite{Ambrosio:2002qq,IceCube:2021eye, Patrizii:2015uea}.}\label{tab:exp-lim}
\end{center}
\end{table}
We estimate $v_{\rm TSM}$ by \cite{Kolb:1990vq}
\begin{align}\label{eq:vM}
v_{\rm TSM}\simeq \mathrm{min}\left[3\times 10^{-3} \left(\frac{m_{\rm TSM}}{10^{16}~\mathrm{GeV}}\right)^{-1/2},1\right] .
\end{align}
\hspace{-1.4mm}In Table \ref{tab:exp-lim} we show the observational upper limits on $\mathcal{F}_{\rm TSM}$ from the MACRO \cite{Ambrosio:2002qq} and IceCube experiments \cite{IceCube:2012khj,IceCube:2021eye}.  The ANTARES telescope \cite{ANTARES:2022zbr} also provides an upper bound on the relativistic monopole flux which is more or less comparable with the Icecube bound. The IceCube experiment also provides \cite{IceCube:2021wvc} a  bound on the flux of low relativistic velocity ($v_{\rm TSM}\simeq 0.1-0.55$) monopoles with masses $\sim 10^{11} - 10^{13}$ GeV.  We have also adopted an observability threshold of $\mathcal{F}_{\rm TSM}$ which is taken to be eight orders of magnitude below the MACRO bound for each value of $G\mu$.  Experiments such as RICE \cite{Hogan:2008sx}, ANITA-II \cite{ANITA-II:2010jck} and Pierre Auger \cite{PierreAuger:2016imq} search for ultra-relativistic monopoles. The most stringent constraint $\mathcal{F}_{\rm TSM}\lesssim 2.5\times 10^{-21} \ \mathrm{cm}^{-2}\mathrm{sec}^{-1}\mathrm{sr}^{-1}$ is provided  by the Pierre Auger Observatory for a Lorentz factor $\gamma = 10^{12}$. From ANITA-II the most strict upper bound on the monopole flux is $10^{-19} \ \mathrm{cm}^{-2}\mathrm{sec}^{-1}\mathrm{sr}^{-1}$ corresponding to $\gamma=10^{10}$. These observational bounds are well below the astrophysical bounds from the survival of the galactic magnetic field, namely, the Parker bound \cite{Parker:1970xv,Turner:1982ag} and its extension in Ref.~\cite{Adams:1993fj} (for recent studies see Refs.~\cite{Kobayashi:2023ryr, Perri:2023ncd}).
\begin{figure}[ht]
\begin{center}
\includegraphics[width = 0.7\textwidth]{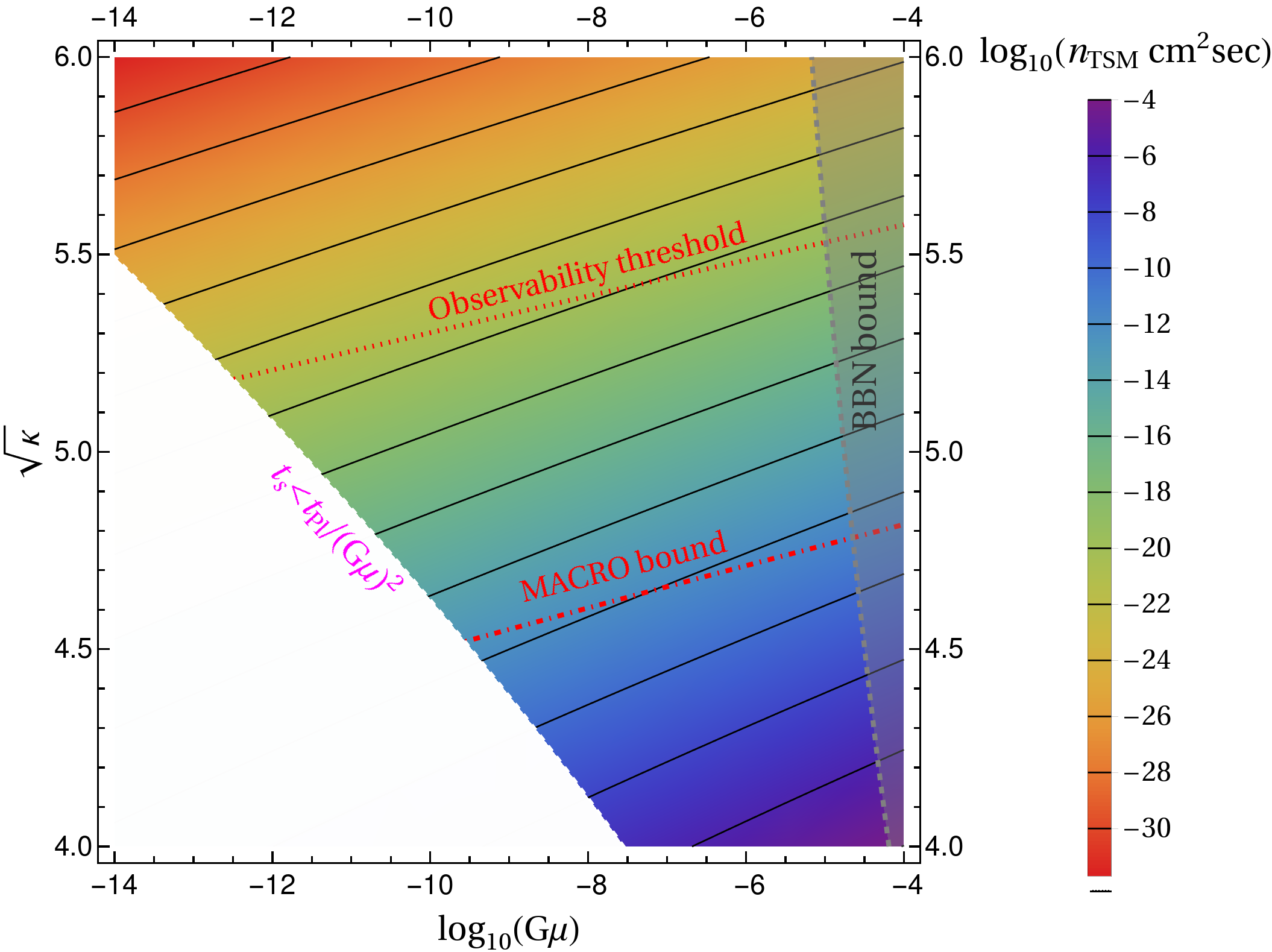}
\end{center}
\caption{Contour plot of the present day TSM number density $n_{\rm TSM}$ in units of $\mathrm{cm}^{-2}\mathrm{sec}^{-1}$ as a function of $G\mu$ and $\sqrt{\kappa}$. The red dot-dashed line dictates the upper limit from the MACRO experiment that $n_{\rm TSM}$ should obey. The lower bound on $\mathcal{F}_{\rm TSM}$ from the adopted observability threshold corresponds to the red dotted line and it is eight orders of magnitude lower than the MACRO bound on $\mathcal{F}_{\rm TSM}$. For $t_s<t_{\rm Pl}/(G\mu)^2$, the evolution of the string network is dominated by friction as it moves in the surrounding plasma and our calculation does not apply. The upper bound on $\Delta N_{\rm eff}$ from the combined BBN and CMB data yields the excluded region shaded in gray (see Section \ref{sec:gws}).}
\label{fig:nMGmukappa_seg}
\end{figure}

It is worth emphasizing that the TSM in our case carries two units of Dirac magnetic charge, whereas the available observational limits that we use are set for singly ($2\pi/e$) charged monopoles. We expect that future data for doubly charged monopoles will not bring about any drastic change in our conclusions, though our bounds on the TSM abundance may somewhat alter. Finally, let us note that our discussion here can be readily extended to $E_6$ grand unification \cite{Gursey:1975ki,Shafi:1978gg,Achiman:1978vg} with the symmetry breaking proceeding via the trinification symmetry group $SU(3)_c\times SU(3)_L\times SU(3)_R$. In this case quantum tunneling would give rise to topologically stable monopoles carrying three units of Dirac magnetic charge \cite{Lazarides:2021tua} from the decay of metastable cosmic strings. An entirely different mechanism for producing such monopoles at an observable level is described in Ref.~\cite{Lazarides:1986rt}. For a discussion of how intermediate mass TSMs in $SO(10)$, produced via the Kibble mechanism during inflation, can be present at an 
observable level, see Refs.~\cite{Lazarides:1984pq, Senoguz:2015lba,  Chakrabortty:2020otp, Lazarides:2021uxv, Maji:2022jzu}.
 For an early study on the relic concentration of monopoles see Refs.~\cite{Zeldovich:1978wj,Preskill:1979zi}.

In Fig.~\ref{fig:nMGmukappa_seg} we depict the present stable monopole number density $n_{\rm TSM}$ in units of $\mathrm{cm}^{-2}\mathrm{sec}^{-1}$ with the red dot-dashed line dictating the upper limit from the MACRO experiment \cite{Ambrosio:2002qq}.
For cosmic times $t<t_{\rm Pl}/(G\mu)^2$, where $t_{\rm Pl}=\sqrt{G}$ is the Planck time, the motion of the strings in the plasma is dominated \cite{Vilenkin:1991zk,Garriga:1993gj} by friction and, thus, the scaling solution cannot be achieved. Consequently, our calculation is valid only for $t_s>t_{\rm Pl}/(G\mu)^2$ which gives the constraint
\begin{align}
\sqrt{\kappa}>(-2.1921\log_{10}(G\mu)-0.585)^{1/2}.
\end{align}
Note that the IceCube bound applies to the relativistic lighter monopoles and the relevant region on the $G\mu$-$\sqrt{\kappa}$ plot lies in the friction dominated era.

 We also note that the EW symmetry is unbroken for all values of $G\mu$ and $\sqrt{\kappa}$ in Fig.~\ref{fig:nMGmukappa_seg} as required for our analysis. 
  In summary, we have shown that, for $G\mu$ between $10^{-9}$ and $10^{-5}$ and for adequately small values of $\sqrt{\kappa}\approx 4.55 - 5.53$, our model predicts the existence of a TSM flux from the decay of metastable strings which should be looked for in the near future. These monopoles have masses $m_{\rm TSM} \approx 3.5 \times 10^{15} - 4.3 \times 10^{17}$ GeV and their mean velocity is expected to vary in the range $4.6\times 10^{-4} - 5 \times 10^{-3}$ \cite{Kolb:1990vq}.
\section{High frequency gravitational waves}
\label{sec:gws}
In this section we briefly discuss the gravitational wave spectrum produced by the metastable string network. The gravitational wave bursts from a cusp is given \cite{Damour:2001bk} by the waveform
\begin{align}
h_c(f,l,z) = g_{1c}\frac{G\mu \, l^{2/3}}{(1+z)^{1/3}r_p(z)}f^{-4/3},
\end{align}
with $g_{1c}\simeq 0.85$ \cite{LIGOScientific:2021nrg}.
The dominant contribution to the radiated gravitational wave power spectrum comes from string loop cusps. The rate of gravitational wave bursts from string loops at frequency $f$ is given by
\begin{align}\label{eq:burst-rate}
\frac{d^2R_c}{dz \, dl} = N_c H_0^{-3}\phi_V(z) \frac{2n(l,t(z))}{l(1+z)}\left( \frac{\theta_m(f,l,z)}{2}\right)^2\Theta(1-\theta_m),
\end{align}
 with an average number of cusps on a loop $N_c=2.13$ \cite{Cui:2019kkd}. Here $H_0$ is the present Hubble parameter, $\theta_m$ the beam opening angle
\begin{align}
\theta_m(f,l,z) = \left[\frac{\sqrt{3}}{4}(1+z)fl\right]^{-1/3},
\end{align}
 and $\phi_V$ is defined as 
\begin{align}
 \phi_V(z) = \frac{4\pi \phi_r^2(z)}{(1+z)^3\mathcal{H}(z)},
\end{align} 
with $\phi_r = \int_0^z \frac{dz'}{\mathcal{H}(z')}$. The Hubble parameter is expressed as
\begin{align}\label{eq:mathcalH}
H(z)\equiv H_0\mathcal{H}(z)=H_0\sqrt{\Omega_{\Lambda,0}+\Omega_{m,0}(1+z)^3+\Omega_{r,0}\mathcal{G}(z)(1+z)^4} \ ,
\end{align}
where $\Omega_{m,0}=0.308$, $\Omega_{r,0}=9.1476\times 10^{-5}$, $\Omega_{\Lambda,0}=1-\Omega_{m,0}-\Omega_{r,0}$ are the present day relic energy fractions of matter, radiation, and dark energy \cite{Planck:2018vyg} respectively and 
\begin{align}
\mathcal{G}(z)=\frac{g_*(z)g_{*S}^{4/3}(0)}{g_*(0)g_{*S}^{4/3}(z)} ,
\end{align}
with $g_*(z)$ and $g_{*S}(z)$ denoting the effective numbers of relativistic degrees of freedom at redshift $z$ for the energy and entropy density respectively \cite{Binetruy:2012ze}.

 The gravitational wave background from string loops at frequency $f$ is given as \cite{Olmez:2010bi, Auclair:2019wcv, Cui:2019kkd, LIGOScientific:2021nrg}
\begin{align}\label{eq:GWs-Omega-cusps}
\Omega_{GW}(f)=\frac{1}{\rho_c}\frac{d\rho_{\rm GW}}{d\ln f} = \frac{4\pi^2}{3H_0^2}f^3\int_{z_*}^{z_F} dz \int dl \, h_
c^2(f,l,z)\frac{d^2R_c}{dz \, dl} \ ,
\end{align}
where $\rho_c$ is the present day critical energy density, $z_F$ is the redshift at time $t_{\rm Pl}/(G\mu)^2$ and $z_*$, which is given by
\begin{align}\label{eq:zstar}
f = \int_{0}^{z_*} dz \int dl \frac{d^2R_c}{dz \, dl},
\end{align}
removes recent resolvable bursts from the background. The limits on $l$ in Eqs.~\eqref{eq:GWs-Omega-cusps} and \eqref{eq:zstar} can be taken from $0$ to the size of the particle horizon at cosmic time $t$. However, various theta functions select the appropriate limits of integration.
\begin{figure}[ht]
\begin{center}
\includegraphics[width = 0.7\textwidth]{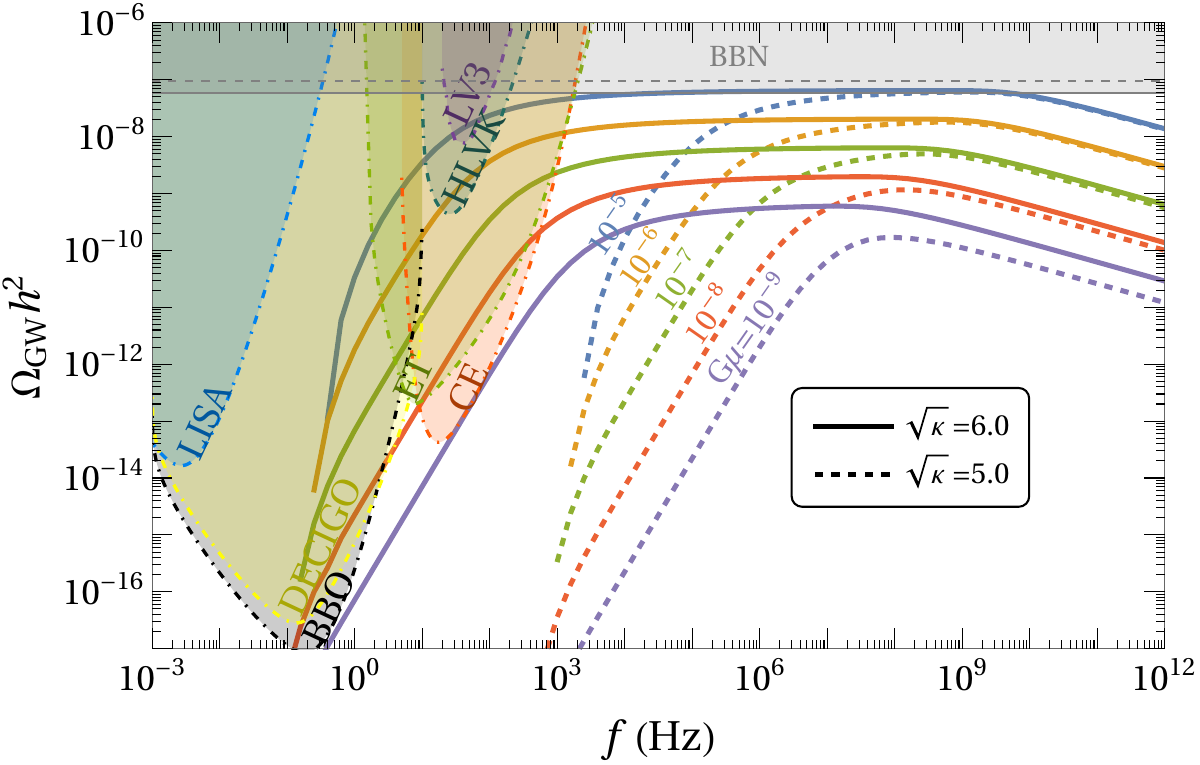}
\end{center}
\caption{Gravitational wave spectra produced by the metastable string network for $\sqrt{\kappa}=5$ (dashed) and $6$ (solid) and $G\mu=10^{-9}-10^{-5}$. We have depicted the power-law integrated sensitivity curves \cite{Thrane:2013oya, Schmitz:2020syl} for planned experiments, namely, HLVK \cite{KAGRA:2013rdx}, ET \cite{Mentasti:2020yyd},  CE \cite{Regimbau:2016ike}, DECIGO \cite{Sato_2017}, BBO \cite{Crowder:2005nr, Corbin:2005ny}, and LISA \cite{Bartolo:2016ami, amaroseoane2017laser}. We have also depicted a gray shaded excluded region from the combined data of BBN and CMB for the nearly scale-invariant plateau region of the spectra for $G\mu=10^{-5}$ with $\sqrt{\kappa}=5$ (dashed boundary) and $6$ (solid boundary).}
\label{fig:GWs}
\end{figure}

 Since the metastability factor $\sqrt{\kappa}$ in our discussion is significantly lower than the values normally chosen to explain the pulsar timing array data, the string network in our case has a correspondingly lower lifetime. It is therefore not surprising that the gravitational wave spectrum is shifted to the higher frequency range.
In Fig.~\ref{fig:GWs} we display the spectrum for two values of the metastability factor, namely $\sqrt{\kappa} = 5$ and $6$ respectively. For $\sqrt{\kappa}$ around $5$, the monopole abundance from the string network decay is in the observable range but the gravitational wave spectrum would be hard to detect in the near future \cite{Aggarwal:2020olq,Bringmann:2023gba,Servant:2023tua}. With $\sqrt{\kappa}\approx 6$, the monopole abundance is quite suppressed, but the gravitational wave spectrum shows a peak in the frequency range that will be tested by HLVK \cite{KAGRA:2013rdx} in the near future, and hopefully by the Einstein Telescope (ET) \cite{Mentasti:2020yyd} and Cosmic Explorer (CE) \cite{Regimbau:2016ike} in the foreseeable future.

The stochastic gravitational wave background contributes to the effective number of relativistic degrees of freedom, which is constrained from the measurement of the effective number of additional dark relativistic degrees of freedom $N_{\rm eff}$ in the  big bang nucleosynthesis (BBN) and cosmic microwave background (CMB) data \cite{Aver:2015iza, Peimbert:2016bdg, Planck:2018vyg}. The ratio of the present day energy density of the gravitational wave background to the critical density is given by \cite{Maggiore:1999vm}
\begin{align}\label{eq:bbn-cmb}
\frac{\rho_{\rm GW}}{\rho_c}\lesssim \Omega_{\gamma,0} \frac{7}{8}\left(\frac{4}{11}\right)^{4/3}\Delta N_{\rm eff},
\end{align}
where $\Delta N_{\rm eff} = N_{\rm eff}-N_{\rm eff}^{\rm SM}$ with $N_{\rm eff}^{\rm SM}=3.044(1)$ \cite{EscuderoAbenza:2020cmq,Akita:2020szl,Froustey:2020mcq,Bennett:2020zkv} being the effective number of neutrinos in the SM and $\Omega_{\gamma,0}\simeq 2.5\times 10^{-5}$ is the present day relic energy fraction in photons. The combined upper bound of BBN and CMB is $\Delta N_{\rm eff}\lesssim 0.22$  \cite{Planck:2018vyg}. Combining Eqs.~\eqref{eq:GWs-Omega-cusps} and \eqref{eq:bbn-cmb}, the constraint from $\Delta N_{\rm eff}$ on the gravitational wave spectra reads:
\begin{align}
\int_{f_{\rm low}}^{f_{\rm high}}\Omega_{GW}(f)d\ln{f} \lesssim \Omega_{\gamma,0} \frac{7}{8}\left(\frac{4}{11}\right)^{4/3}\Delta N_{\rm eff} .
\end{align}
In our case, the network disappears in the pre-BBN era and we can, therefore, take the lower limit of the integral to be 
\begin{align}
f_{\rm low} \sim 10^{-9}~\mathrm{Hz}~\left(\frac{G\mu}{10^{-7}}\right)^{-1/2}e^{-\frac{\pi}{4}(\kappa-64)},
\end{align}
which is one order of magnitude smaller than the lower cut-off frequency of the scale-invariant plateau region of the gravitational wave spectra  \cite{Buchmuller:2021mbb}.
For the upper bound of the integral we take $f_{\rm high}=10^{12}$ Hz without loss of generality (see Fig.~\ref{fig:GWs}).

The excluded parameter space from the upper bound on $\Delta N_{\rm eff}$ is depicted by the gray shaded region in Fig.~\ref{fig:nMGmukappa_seg}. In  Fig.~\ref{fig:GWs}, we have shown a gray shaded excluded region for the nearly scale-invariant plateau ($\Omega_{\rm GW}\propto f^0$) of the spectra for $G\mu=10^{-5}$ with $\sqrt{\kappa}=5$ and $6$.

\section{Monopole production in friction dominated era}
\label{sec:fric-era}
\begin{figure}[ht]
\begin{center}
\includegraphics[width = 0.7\textwidth]{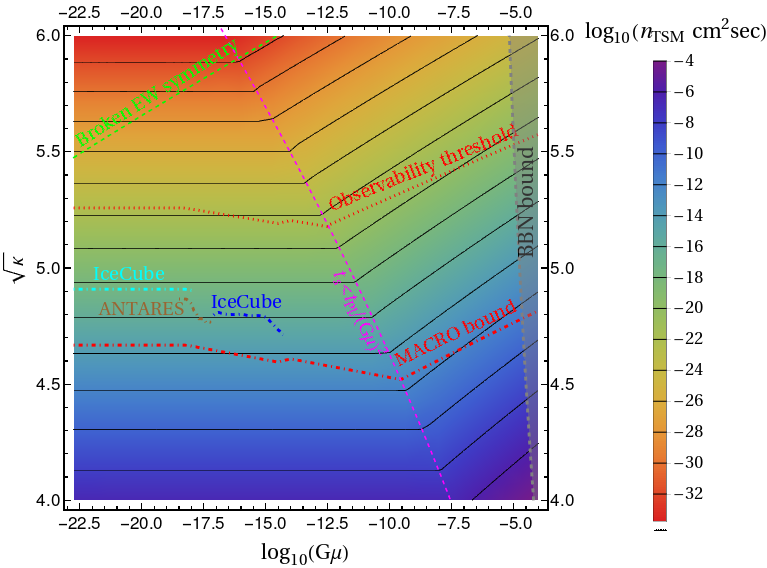}
\end{center}
\caption{Contour plot of the present day TSM number density $n_{\rm TSM}$, in units of $\mathrm{cm}^{-2}\mathrm{sec}^{-1}$, as a function of $G\mu$ and $\sqrt{\kappa}$ for the string networks that produce sub-horizon segments before and after early friction domination. In the region above the green dashed line the EW symmetry is broken. The cyan and blue dot-dashed lines display the IceCube bounds for $v_{\rm TSM}\simeq 0.8-1$ and $v_{\rm TSM}\simeq 0.1-0.55$ respectively. We also show a comparable upper bound from ANTARES with the brown dot-dashed line for $v_{\rm TSM}\simeq 0.55-0.995$. The red dot-dashed line displays the upper limit on $n_{\rm TSM}$ from the MACRO experiment. The red dotted line depicts the adopted observability threshold. We show the variation of $n_{\rm TSM}$ with a rainbow code.}\label{fig:nTSM_fric}
\end{figure}
In this section we provide an estimate for the monopole number density from a metastable string network which spontaneously decays via quantum tunneling and produces sub-horizon segments before $t_F = t_{\rm Pl}/(G\mu)^2$, that is $t_s < t_F$. These segments do not oscillate in the friction dominated era and remain frozen until the friction domination ends at time $t_F$.
 We expect that the string network has a total length of order $p\, t_F$ within the particle horizon at time $t_F$ , where $p$ is a geometric factor to be estimated below. The length of a string segment at time $t_F$ is given by \cite{Leblond:2009fq}
\begin{align}
l \sim \frac{1}{\Gamma_dt_F}=\frac{t_s^2}{t_F},
\end{align}
and the total number of segments within the horizon at time $t_F$ will be $p\,t_F/l$.
The segments with length $1/(\Gamma_dt_F)$ decay very quickly at the end of friction domination at $t_F$. Therefore, the present day number density of TSM is given by
\begin{align}
n_{\rm TSM}\sim \frac{p}{t_s^2t_F}\frac{1}{(1+z_F)^3} ,
\end{align}
where $z_F$ is the redshift at time $t_F$. The monopole number densities from the networks with sub-horizon segments before ($t_s < t_F$) and after ($t_s > t_F$) the early friction domination match at $t_F$ for $p \sim 10$.

Fig.~\ref{fig:nTSM_fric} displays the monopole number densities from the metastable string networks that decay before and after the early friction 
domination. The upper limit on the relativistic TSM number density from the IceCube experiment corresponds to an almost horizontal line segment at $\sqrt{\kappa} \simeq 4.9$ and $G\mu \lesssim 5.6 \times 10^{-18}$. On this line segment $v_{\rm TSM} \gtrsim 0.8$, $m_{\rm TSM}$ lies between $3.4 \times 10^{8}$ GeV and $1.4 \times 10^{11}$ GeV, and the symmetry breaking scale associated with the metastable strings is smaller than about $1.6 \times 10^{10}$ GeV. Note that $v_{\rm TSM}$ is calculated from Eq.~\eqref{eq:vM}. We expect that, in the foreseeable future, the IceCube experiment will be able to detect fluxes of relativistic monopoles higher than its present limit. For smaller values of $G\mu$ the monopoles become ultra-relativistic and 
we hope that their flux will be detectable in the foreseeable future 
at the Pierre Auger and ANITA experiments. Note that the present 
bounds from these experiments hold for monopole masses lower than 
about $10^{7}$ GeV.

\section{Conclusions}
\label{sec:conc}
We have explored a novel cosmological scenario for producing topologically stable magnetic monopoles in the early universe from the decay of metastable cosmic strings after the end of the friction dominated era. We find that an experimentally observable flux of superheavy ($\sim 10^{15}-10^{17}$ GeV) monopoles, not far below the MACRO bound, is realized in the framework of $SO(10)$ grand unification for dimensionless string tension parameter values $G\mu \approx 10^{-9} -10^{-5}$ and  the metastability factor $\sqrt{\kappa}$ around 5. We also explore smaller $G \mu$ values, of order $10^{-22}-10^{-10}$, that give monopole masses of order $10^{8} - 10^{14}$ GeV with a flux that should be accessible at experiments including IceCube, Pierre Auger, ANITA, and future observatories such as KM3NeT \cite{Spurio:2012mh}.
The gravitational wave emission from the metastable cosmic strings with $G\mu\approx 10^{-9}-10^{-5}$ and $\sqrt{\kappa} \approx 6$ lies in a wide frequency range accessible at HLVK and future detectors including ET and CE.
An $E_6$ based extension of our discussion here will give rise to topologically stable monopoles carrying three quanta of Dirac magnetic charge from the quantum mechanical decay of metastable strings.
The discovery of magnetic monopoles would have far reaching consequences for searches of new physics beyond the Standard Model.
 
\section{Acknowledgment}
This work is supported by the National Research Foundation of Korea grant: 2022R1A4A5030362 by the Korean government, IBS under the project code: IBS-R018-D3 (R.M.) and the Hellenic Foundation for Research and Innovation (H.F.R.I.) under the ``First Call for H.F.R.I. Research Projects to support Faculty Members and Researchers and the procurement of high-cost research equipment grant" (Project Number: 2251) (G.L. and Q.S.).  R.M. would like to thank Wan-Il Park for many illuminating discussions and acknowledge the warm hospitality at the Indian Association for the Cultivation of Science, Kolkata, where a part of the research was carried out.  Q.S. thanks Amit Tiwari for discussions.

\bibliographystyle{JHEP}
\bibliography{strings_mon_MSS_6}

\end{document}